# SAKR: Enhancing Retrieval-Augmented Generation via Streaming Algorithm and K-Means Clustering

Haoyu Kang[1*], Yuzhou Zhu[2*], Yukun Zhong[3*], Ke Wang[4] and Ping Zhong[1†]

[1] Central South University, Changsha, China
[2] Dalian University of Technology, Dalian, China
[3] Nanjing University, Suzhou, China
[4] Xidian University, Xi'an, China

**Abstract.** Retrieval-augmented generation (RAG) has achieved significant success in information retrieval to assist large language models (LLMs) in answering questions from unseen documents by building an external knowledge base. However, it faces significant challenges, including high memory consumption—due to a large external database—and the inability to update this index in real time when handling large-scale streaming data. To reduce the memory required for building the database and maintain accuracy simultaneously, we proposed a novel approach that integrates a streaming algorithm with k-means clustering into RAG. Our approach applies a streaming algorithm to dynamic index updates and reduces memory consumption. Additionally, the k-means clustering algorithm that groups similar documents is applied to reduce query time. We conducted comparative experiments on RAG with streaming algorithm and k-means clustering (SAKR), and the results indicated that SAKR outperforms traditional RAG in both accuracy and memory efficiency, particularly for large-scale streaming data, with an average accuracy of 0.640 and 10% memory cost.

## 1    Introduction

Retrieval-augmented generation (RAG) is a mechanism that enhances the performance of large language models (LLMs). RAG establishes an external knowledge database, enabling LLMs to generate more accurate, time-sensitive answers, particularly for queries involving unseen or real-time information [1].

* These authors contribute equally to this work
† Corresponding Author (e-mail: ping.zhong@csu.edu.cn)

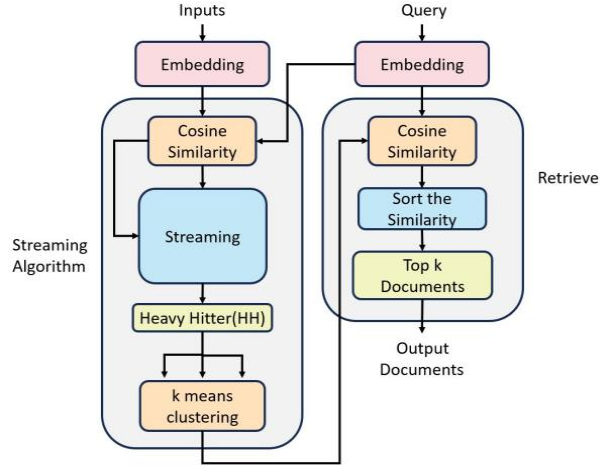

Fig. 1. Schema of our method

However, traditional RAG has limitations in handling massive streaming data and costs a huge amount of memory and efficiency. RAG performs with low accuracy in dealing with unseen data [34]. Since streaming data consists of unseen data, RAG shows poor accuracy when dealing with streaming data that changes frequently, such as new articles. RAG also costs a huge memory due to building an external knowledge database that stores large amounts of documents [2]. Additionally, RAG must handle many queries in real-world scenarios, and the query time should not be too long. As the number of relevant documents grows, the efficiency of traditional RAG decreases significantly [34].

To solve these two problems, we implement a transformed heavy hitters streaming algorithm within RAG to reduce memory usage by filtering out documents unlikely to be retrieved. Traditional RAG costs huge amounts of memory because it builds a large database to store documents. Therefore, if we want to optimize the memory, it is evident that we can try to reduce the size of the big database. According to previous research [10], streaming algorithms effectively reduce memory, so we came up with an idea to apply a streaming algorithm to RAG to see if it works. Heavy hitters is a streaming algorithm that filters and retains data occurring more frequently. The streaming algorithm in our method is transformed heavy hitters algorithm. Our method filters the databases, leaving documents more likely to result from a query. Implementing the streaming algorithm can also update the index database to improve the low precision of traditional RAG when dealing with massive data that is frequently changing. Then we combined the streaming algorithm with the k-means cluster algorithm to enhance performance further. We used the k-means cluster algorithm to cluster similar documents together in advance after selecting the streaming algorithm, and it optimizes the query time because it can retrieve documents from K clusters rather than the whole index database while inputting a new query.

We conducted experiments on a news dataset with constantly updated messages and calculated the precision, recall rate, and F1 score of the proposed SAKR approach to compare the accuracy of SAKR with Naive RAG. Our experiments' results showed that implementing the k-means clustering algorithm and streaming algorithm can effectively save 90% of memory and exhibit greater accuracy when the number of queries is large. Our results demonstrated that the streaming algorithm optimized the memory space while maintaining relatively high accuracy. Additionally, the ablation study highlights the significance of each component in our approach, underscoring their contributions to performance improvements.

## 2    Related Work

### 2.1    The evolution of RAG and its problems

RAG was first proposed in the article "Retrieval-Augmented Generation for Knowledge-Intensive NLP Tasks" [2]. This article illustrated the limited ability of LLMs to access and precisely manipulate knowledge, so LLMs' performance on knowledge-intensive tasks lags task-specific architecture.
Plus, pre-trained LLMs cannot easily expand or revise their memory, and they encounter challenges like hallucination, outdated knowledge, and non-transparent, untraceable reasoning processes [1]. So, this information retrieval mechanism, Retrieval-augmented generation, applies to LLMs to solve the problems that pre-trained LLMs face. This model, Naive RAG, builds an external knowledge database and retrieves relevant documents to help LLMs. However, it has some drawbacks, such as low query precision, missing important information, etc.

The emergence of Advanced RAG was aimed at improving its shortcomings [3]. Advanced RAG enhances its indexing techniques by employing a sliding window approach, fine-grained segmentation, and metadata integration. Furthermore, it integrates multiple optimization methods to streamline the retrieval process. Each community will generate a community answer and then incorporate it into a global answer. They improved the RAG's performance in terms of the comprehensiveness and diversity of answers.

Microsoft found that Naive RAG is not good at extracting global answers and fails on query-focused summarization tasks (QFS). So, they proposed an approach that builds the knowledge graph based on the extracted entities and relationships in the sentence and then divides the graph into communities using the Leiden algorithm [4]. Each community will generate and incorporate a community answer into a global answer. They improved the RAG's performance in terms of the comprehensiveness and diversity of answers. Comprehensive surveys have summarized these advances—covering retriever–generator coupling, pipeline segmentation, multimodal extensions [35]—and, from a pre-/post-retrieval perspective and outlined evaluation methods[36]. Inspired by previous research, we realized that the proposal of traditional RAG is significant, but there are still a lot of things to improve. What we want to do is to save the huge memory cost and the huge amount of time spent on each query of relevant documents.

## 2.2 Streaming Algorithms

The streaming algorithm was first regularized and popularized in the paper "The Space Complexity of Approximating the Frequency Moments" [6]. Streaming algorithms process the streaming data where the input is a boundless and ordered sequence of items, which can usually be handled in a few passes and only once in most cases [7]. However, in most cases, the true class label of streaming data is probably unavailable, and there is no prior information on the number of classes, so the clustering algorithm is a good option to cluster and define the class label of the streaming data [8]. Generally, the streaming algorithm adapts traditional algorithms to function effectively with streaming data regarding limitations such as limited memory, single-pass processing, and real-time analysis. The Streaming clustering methods can be classified into five categories: partitioning, hierarchical, density-based, grid-based, and model-based [10].

**Partitioning-based methods:** Streaming data is categorized into a pre-defined number of clusters according to the distance or similarity to the center of the clusters. Specifically, Incremental k-means [12], StreamKM++ [13], and SWClustering [14] belong to this method.

**Hierarchical-based methods:** These algorithms apply Binary Tree and can be sorted into two categories: Aggregate and Split algorithms. Aggregate algorithms assume that every item forms a cluster initially and gradually merge the items (subclusters) to form new larger clusters. However, split algorithms consider that all the streaming data is comprised of a cluster and progressively split it up. Birch [15], CluStream [11], and HUEStream [16] are typical hierarchical-based methods.

**Density-based methods:** These methods summarise input data into many microclusters, which are then updated based on density accessibility and connectivity. DenStream [17], OPTICS-Stream [18], and D-Stream [19] are some popular methods of this kind.

**Grid-based methods:** The whole workspace is divided into numerous grid units, each containing a data item. Then we cluster the grid units according to their density. MR-Stream [20] and CellTree [21] are designed based on the preferred methods.

**Model-based methods:** These methods focus on finding the data distribution model that best fits the input streaming data, such as SWEM [22] and GCPSOM [23].

Identifying Heavy Hitters, often called finding the top k popular or frequent items, is a well-researched and important problem in streaming data [5]. The two main kinds of Heavy Hitters algorithms are methods based on the counter and those based on the sketch. The former methods keep track of the frequent elements by maintaining an array of counters that directly calculate the numbers. Misra-Gries [24], Lossy Counting [25], and Space Saving [26] belong to the algorithms based on counters. Methods based on sketches apply probabilistic data structures such as Hash Functions and Bitmaps to describe the streaming data, providing frequency estimates with less memory consumption.CountSketch [27], CountMinSketch [28], Multi-stage Bloom Filters [29], and Sketch-guided Sampling [30] are part of these methods. After realizing the power of the streaming algorithm, we want to use the heavy hitter algorithm to build and update the index database to solve the problem of RAG consuming a lot of memory.

## 3    Method and Algorithm

### 3.1    Dataset

The Dataset that we use is at Kaggle [9]. It contains more than sixteen thousand pieces of news and comments and 23 features from January 1, 2020, to December 31, 2020. We selected three features to help us retrieve relevant documents: headlines, keywords, and abstracts. We combined these three features into one feature and then embedded the new combined feature into a vector. Since human annotation is one of the solutions for evaluating RAG [33], we manually labeled the dataset to calculate the accuracy of our input query. We marked the documents that are related to the query we entered as 1, and we marked the rest as 0.

### 3.2    Algorithm

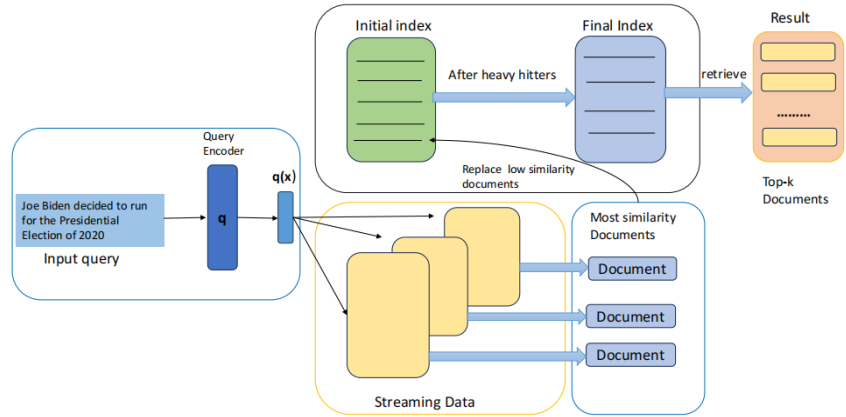

Fig. 2. Workflow of our method

Figure 2 is a workflow that describes how our algorithm works. In the preparation of our algorithm, we embed the query into a vector, split the input data into chunks, and set up a new data structure to store the documents, which only costs 10% of the memory of the Naive RAG database.

In the initialization of our algorithm, it reads the streaming data to fill the new data structure and sorts the data structure for the convenience of replacing the document with the lowest similarity each time in the following chunks. Then, it reads the streaming data continuously and implements the transformed heavy hitter algorithm: For every input chunk, we calculate the highest cosine similarity in this chunk and use it to replace the document with the lowest similarity in our new database. When the streaming data ends, we get the final database that stores all the documents filtered by the transformed heavy hitter algorithm.

$$\text{CosSim}(\mathcal{Q}, \mathcal{D}_i) = \frac{\mathcal{Q} \cdot \mathcal{D}_i}{\|\mathcal{Q}\|\|\mathcal{D}_i\|}$$

$$p_{\text{HH}}(d_i|q) = \sigma\left(\alpha \cdot \text{CosSim}(\mathcal{Q}, \mathcal{D}_i) - \beta\right)$$

where:

- $\mathcal{D}_i$ is the result of embedded data
- $\mathcal{Q}$ is the result of the embedded query
- $\sigma(x) = \frac{1}{1+e^{-x}}$ is the sigmoid function, and $\alpha$ and $\beta$ are hyperparameters.

Algorithm 1 is the pseudo-code of our method.

**Algorithm 1** Streaming Algorithm (Heavy Hitter) Implementing in RAG to Update Its Index

**Require:** streaming_data, size_index: $n$
**Ensure:** top-$k$ vectors
0: **Initialize:**
0: **for** $i \leftarrow 0$ to $n-1$ **do**
0:   index[$i$] $\leftarrow$ streaming_data
0: **end for**
0: sort index
0: **while** not empty_stream_data() **do**
0:   sort documents in stream_data
0:   output the highest similarity documents doc[$i$]
0:   index[$n$] $\leftarrow$ doc[$i$]
0:   $n \leftarrow n - 1$
0: **end while**=0

The above pseudo-code describes how Streaming RAG works. However, a key issue arises when handling new queries and retrieving relevant documents: the need to sort the similarity of all documents. This process is computationally expensive and time-consuming. To address this problem, we combine the k-means clustering algorithm with streaming RAG to group n documents into m clusters. This clustering step enables faster retrieval of relevant documents by narrowing the search space, thus reducing the time required for similarity calculations during query processing.

Where:

n is the number of all the documents in the database.

m is the number of clustered classes.

n >> m

The following formulas describe the clustering process and retrieval of the top-k documents:

$$\text{Euclidean Distance}(\mathcal{Q}, \mathcal{D}_i) = \sqrt{2(1 - \text{CosSim}(\mathcal{Q}, \mathcal{D}_i))}$$

$$\mathcal{C}_j = \frac{1}{|\mathcal{H}_j|} \sum_{i \in \mathcal{H}_j} \mathcal{D}_i$$

$$p_{\text{top-k}}(d_i|q) = \sum_{j \in \text{top-k}(\text{CosSim}(\mathcal{Q},\mathcal{C}_j))} p_{\text{HH}}(d_i|q) \cdot p_{\text{kmeans}}(d_i|\mathcal{C}_j)$$

$$p_{\text{RAG-Token}}(y|x) \approx \prod_{i=1}^{N} \sum_{z_i \in \text{top-k}(p(\cdot|x))} p_\eta(z_i|x) p_\theta(y_i|x, z_i, y_{1:i-1})$$

where:
- $\mathcal{H}_j$ is the index set of the $j$-th cluster

In practice, you will deal with many queries, and then the time cost of clustering in advance can be ignored. The time complexity of searching of traditional RAG model is O(N · nlogn), but the Streaming algorithm and K-means RAG are O(N · mlogm) where:

N is the number of queries.

Then, to retrieve an article, we only need to search from these m classes rather than n documents. Table 1 describes the time complexity of two methods.

TABLE I
COMPARISON ON TIME COMPLEXITY OF TWO METHODS

| Method | Time complexity |
|---|---|
| SAKR | $O(mlogm)$ |
| RAG | $O(nlogn)$ |

## 4   Experiments and Results

We conducted three experiments to evaluate the performance of our method. All experiments were based on our dataset and used the input query: "Biden decided to run for the Presidential Election of 2020." The goal of these experiments was to assess the memory and accuracy performance of SAKR (Streaming algorithm and K-means RAG), compare the effects of the k-means clustering algorithm with and without a streaming algorithm, and explore the influence of different cluster sizes on accuracy. Additionally, we aimed to investigate the trade-off between memory usage and accuracy.

### 4.1 Performance Comparison

The prediction accuracy results are shown in Table 2. When dealing with large-scale streaming data or when the number of retrieved documents is high, SAKR ranked first in all evaluation metrics, reflecting a significant improvement in performance over Naive RAG. Additionally, SAKR only occupies 10% memory of Naive RAG.

TABLE II
PERFORMANCE COMPARISON (K=50)

| Metrics | Precision | Recall | F1 | Memory Ratio(%) |
|---|---|---|---|---|
| Naive RAG | 0.720 | 1.000 | 0.840 | 100% |
| SAKR | 0.840 | 1.000 | 0.910 | 10% |

### 4.2 Comparative experiments on clustering effects

To illustrate the rationale behind implementing the streaming algorithm before k-means clustering, we experimented to compare the effects of the k-means clustering algorithm in two scenarios: clustering before the streaming algorithm and clustering after the streaming algorithm. As shown in Figure 4, it is evident that the clustering results after applying the streaming algorithm are significantly better than those obtained before. As shown in Table 3, this conclusion is further supported by the silhouette coefficients [33] calculated for both clustering methods.

TABLE III
SILHOUETTE COEFFICIENT OF TWO METHODS

| Method | silhouette coefficient |
|---|---|
| K-means RAG | 0.2405110 |
| Streaming and K-means RAG | 0.4879005 |

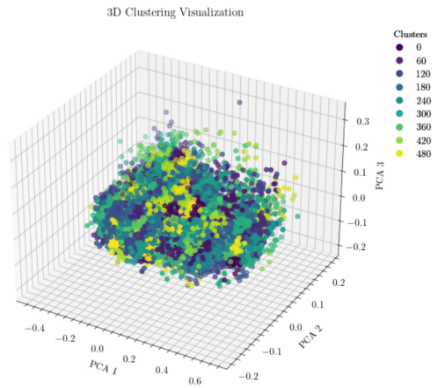 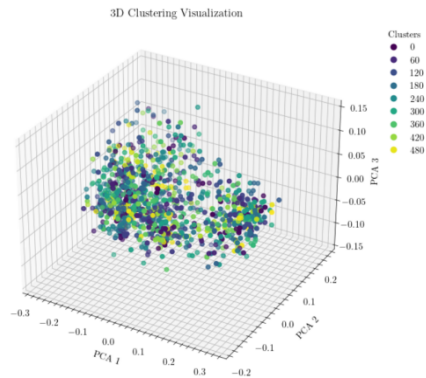

Fig. 3. Clustering results before streaming algorithm    Fig. 4. Clustering results after streaming algorithm

### 4.3 The effect of a different number of clusters on accuracy

This experiment explored the relationship between retrieval accuracy and the number of clusters. Figure 5 illustrates how accuracy varies with the number of clusters. As shown, the retrieval accuracy increases as the number of clusters grows, but the rate of improvement slows down, and the computational cost of k-means clustering also increases. Additionally, we can observe that the accuracy improves only marginally once the number of clusters exceeds 800. Therefore, we chose several clusters between 600 and 800 to better balance the accuracy and time efficiency.

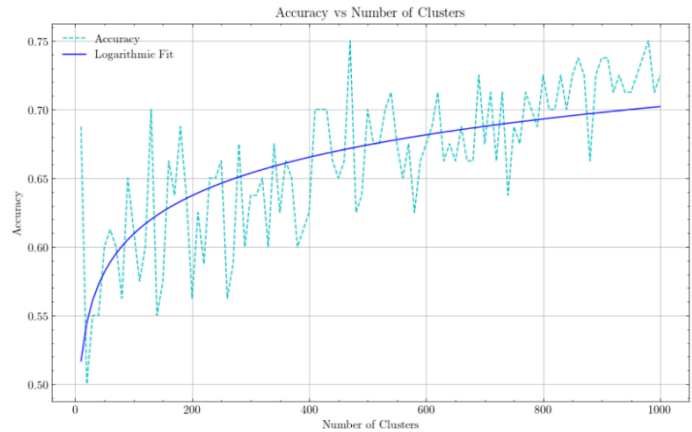

Fig. 5. Variation of Accuracy with the Number of Clusters

### 4.4 Ablation Study

We conducted ablation studies on SAKR. Firstly, we omitted the streaming algorithm to evaluate the effectiveness of clustering. As shown in Table 4, it is evident that compared to the method using a streaming algorithm, SAKR significantly increases the accuracy for each value of K. Subsequently, we compared the prediction accuracy of SAKR with the approach that does not use the K-means clustering algorithm, retaining only the streaming algorithm and RAG. As shown in Figure 5, the latter achieves slightly higher accuracy than SAKR. However, as discussed in the Methods section, implementing the K-means clustering algorithm helps avoid the need to sort documents for every query.

TABLE IV
THE RESULT OF ABLATION EXPERIMENTS

| K | 1 | 10 | 20 | 30 | 40 | 50 | 60 | 70 | 80 | 90 | 100 | Memory Ratio(%) |
|---|---|---|---|---|---|---|---|---|---|---|---|---|
| SAKR - Clustering | 1.000 | 0.800 | 0.800 | 0.767 | 0.775 | 0.720 | 0.750 | 0.757 | 0.725 | 0.700 | 0.680 | 10% |
| SAKR-streaming | 0.800 | 0.640 | 0.470 | 0.360 | 0.345 | 0.312 | 0.283 | 0.266 | 0.258 | 0.251 | 0.234 | 100% |
| SAKR | 1.000 | 0.700 | 0.550 | 0.667 | 0.700 | 0.680 | 0.700 | 0.657 | 0.638 | 0.633 | 0.640 | 10% |

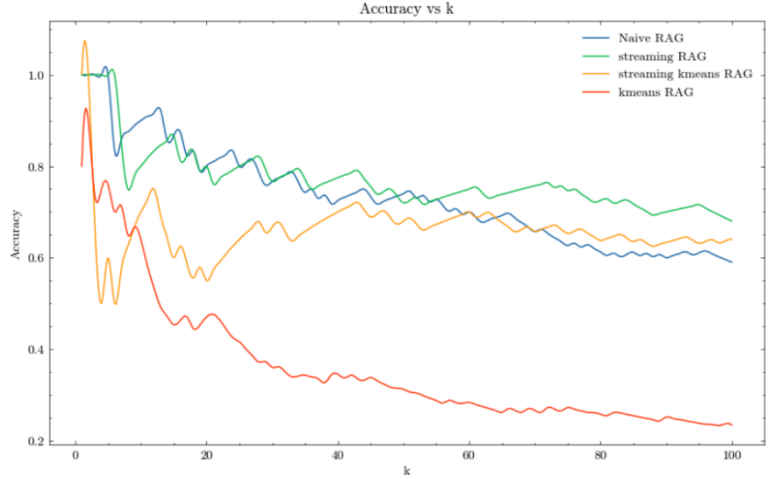

Fig. 6. Comparison of the accuracy of four algorithms for different values of k

## 5      Conclusion

In this paper, we introduce a novel approach that integrates streaming algorithms with k-means clustering within the RAG framework. Our method achieves a 90% reduction in memory consumption and enhances accuracy when handling streaming data. Furthermore, by narrowing the query scope through clustering, we significantly reduce query times. These improvements make our method more resource-efficient and faster compared to traditional RAG, particularly in large-scale streaming data scenarios. By plotting the accuracy variation curve against the number of clusters, we concluded that the optimal number of clusters lies between 600 and 800, corresponding to approximately 3.75% to 5% of the total dataset size, providing a trade-off between query time and accuracy.

In addition to issues such as excessive memory usage and decreased query accuracy when dealing with streaming data, traditional RAGs also suffer from poor performance in extracting global information. Our method has not yet addressed this problem. To tackle this issue, Microsoft has proposed a GraphRAG model for better extracting global information and the main idea. However, it consumes significant memory due to storing graph data structures. In future work, combining streaming algorithms with Graph RAG holds great potential [32] for solving the huge memory consumption that Graph RAG takes to build the knowledge graph.

## Acknowledgement:

This work was supported in part by the National Natural Science Foundation of China under 62172443.